\documentclass[preprint,12pt]{elsarticle}

\usepackage{amssymb}

\usepackage{graphicx}
\usepackage{color}
\usepackage{bm}

\journal{Phys. Lett. B}

\begin{document}

\begin{frontmatter}



\title{ Velocity Dependence of Charmonium Dissociation Temperature in High-Energy Nuclear Collisions }

\author[thu,frankfurt]{Yunpeng Liu}
\author[lbnl,ccnu]{Nu Xu}
\author[thu]{Pengfei Zhuang}

\address[thu]{Physics Department, Tsinghua University, Beijing 100084, China}
\address[frankfurt]{Institut f\"ur Theoretische Physik, J. W. Goethe-Universit\"at, Frankfurt, Germany}
\address[lbnl]{Nuclear Science Division, Lawrence Berkeley National Laboratory, Berkeley, California 94720, USA}
\address[ccnu]{College of Physical Science and Technology, Central China Normal University, Wuhan, China}
\begin{abstract}
In high-energy nuclear collisions, heavy quark potential at finite
temperature controls the quarkonium suppression. Including the
relaxation of the medium induced by the relative velocity between
quarkonia and the deconfined expanding matter, the Debye screening
is reduced and the quarkonium dissociation takes place at a higher
temperature. As a consequence of the velocity dependent dissociation
temperature, the quarkonium suppression at high transverse momentum
is significantly weakened in high energy nuclear collisions at RHIC
and LHC.
\end{abstract}


\end{frontmatter}



Heavy quarkonia $J/\psi$ and $\Upsilon$ are tightly bound hadronic
states. Their dissociation temperature $T_d$ is, in general, higher than
the critical temperature $T_c$ for the deconfinement phase
transition~\cite{Asakawa:2003re} in high-energy nuclear
collisions~\cite{Gonin:1996wn, Adare:2006ns, Pillot:2011zg}.
Therefore, the measured cross sections of quarkonia carry the
information of the early stage hot and dense medium. They have long
been considered as a signature of the formation of the new state of
matter, the so-called quark-gluon plasma~\cite{Matsui:1986dk,
Blaizot:1996nq}.

The quarkonium dissociation in a static deconfined quark matter is
generally described in terms of the screening effect. The heavy quark potential, which is normally taken as the Cornell
form~\cite{Eichten:1978tg} and can be calculated through a non-relativistic quantum
chromodynamic potential~\cite{Brambilla:1999xf} and lattice
simulations~\cite{Chen:1998ct}, is reduced to a Yukawa-like
potential due to the Debye screening. When the screening radius
becomes smaller than the quarkonium size, the bound state
dissociates. Substituting the screened potential, extracted from
lattice simulations~\cite{Kaczmarek:2004gv, Digal:2005ht}, into the
Schr\"odinger equation for the wave function of the quarkonium
state, one obtains the dissociation temperature that corresponds to
the zero binging energy and infinite size of the di-quark system~\cite{Satz:2005hx, Wong:2004zr}. For charmonia, while the
excited states $\chi_c$ and $\psi'$ start to dissociate already
around $T_c$,  the calculated dissociation temperature for the
ground state $J/\psi$ is much higher than the critical
temperature~\cite{Satz:2005hx}.

The quarkonia produced in relativistic heavy ion collisions are,
however, not at rest in the medium. There exists a relative velocity
between the quarkonia and the expanding medium. The question is what
is the velocity dependence of the heavy quark potential at finite
temperature~\cite{Liu:2006nn, Escobedo:2011ie, Aarts:2012ka}. The screening effect is due to the rearrangement of the
charged particles when a pair of heavy quarks (source) is present in
the medium. For a moving source, it will take a longer time for the
source to interact with the medium, comparing with that of a
stationary source. This `delay' of the response reduces the
screening charges around the source and thus weakens the screening
effect. In relativistic heavy ion collisions, the average
transverse momentum of the initially produced $J/\psi$s is about $2\textrm{ GeV}$\ at RHIC energy~\cite{Adare:2006kf} and $3\textrm{ GeV}$\ at LHC energy.~\cite{Abelev:2012kr}, corresponding to an averaged relative velocity above $0.5c$. A significant
modification of the Debye screening is expected for such fast moving
$J/\psi$s, especially for those produced in higher transverse
momentum region~\cite{Chu:1988wh,Mustafa:2004hf}. In this Letter, we
study the velocity dependence of the heavy quark potential and the
quarkonium dissociation temperature in a transport approach. The
velocity induced effects on charmonium suppression at both
Relativistic Heavy Ion Collider (RHIC) and Large Hadron Collider
(LHC) will be discussed. In the following calculations we take the
speed of light $c=1$.

For a static source located at ${\bf r}=0$, the ambient charge
density $\rho_0({\bf r})$ is modified by the screening potential
$V_0({\bf r})$ at finite temperature $T$~\cite{debye},
\begin{equation}
\rho_0({\bf r})=\sum_i q_i f_i e^{-q_iV_0({\bf r})/T}\approx -\sum_i
 ( q_i^2 f_i/T ) V_0({\bf r}),
 \label{debye}
\end{equation}
where $q_i$ is the charge of the particles of species $i$, and
$f_i$ is the initial particle density without the source. The neutrality condition for the total charge has been
considered here. The solution of (\ref{debye}) with the assumption of small $V$ gives the Debye screening
of the potential. At large distance, the potential is weak, so that the approximation in (\ref{debye}) is appropriate, while at small distance, the solution of (\ref{debye}) means a small correction to the original potential, as the lattice simulations indicated~\cite{Kaczmarek:2004gv,Digal:2005ht}.

For a source moving with velocity ${\bm \upsilon}$\ with respect to the medium, the non-equilibrium charge density $\rho({\bf r},t)$  in the
source-rest frame satisfies the transport equation in the relaxation
time approximation,
\begin{equation}
\partial_t \rho -{\bm \upsilon}\cdot\nabla \rho = -\left(\rho-\rho_0\right)/\tau,
\label{transport}
\end{equation}
where $\tau$ is the relaxation time of the medium. Taking the limit
$t\rightarrow\infty$, the final distribution $\rho_f({\bf r}, {\bf
L})\equiv \lim_{t\rightarrow \infty} \rho({\bf r},t)$ becomes stable
and is characterized by the equation
\begin{equation}
{\bf L}\cdot\nabla\rho_f =\rho_f-\rho_0,
\label{stable}
\end{equation}
where we have introduced the relaxation length defined as ${\bf L}
\equiv {\bm \upsilon}\tau$ which controls the velocity dependence of the
Debye screening. Since the screening charge distribution is
proportional to the screening potential, see equation (\ref{debye}),
the potentials $V_0$ and $V$ corresponding to a stationary and
moving source, respectively, satisfy the same equation
(\ref{stable}) which can be solved analytically,
\begin{equation}
   V({\bf r},{\bf L})=\int_0^\infty V_0({\bf r}+\lambda {\bf L})e^{-\lambda}d\lambda.
\label{solution2}
\end{equation}
It is obvious that for a static source with $L=0$ we have $V({\bf
r}, {\bf 0})=V_0({\bf r})$.

With the potentials $V_0$ and $V$, the screening radius $r_d$ can be
expressed as
\begin{equation}
   r_d({\bf L})= {1\over 2}{\int d^3{\bf r}\ r \rho_f({\bf r},{\bf L})\over \int
   d^3{\bf r}\ \rho_f({\bf r},{\bf L})}={1\over 2}{\int d^3{\bf r}\ rV({\bf r},{\bf L})\over \int d^3{\bf r}\ V_0({\bf
   r})}.
\label{eqrd}
\end{equation}
For the second equality, we have used the total charge conservation $\int d^3{\bf r} \rho_f({\bf r}, {\bf L})=\int d^3{\bf r} \rho_0({\bf r})$ for any ${\bf L}$ which is guaranteed by integrating Eq.(\ref{stable}) over the whole coordinate space. The screening radius $r_d({\bf L})$ is in general an angle dependent function. However, for a spherically symmetric potential $V_0(r)$, the integration over the angles in the numerator of Eq.(\ref{eqrd}) can be analytically done, and the averaged screening radius can be effectively expressed as $r_d(L)=\int dr r^3 V(r,L)/(2\int dr r^2 V_0(r))$ with the factorized averaged potential
\begin{equation}
V(r,L)=V_0(r) W(r/L),
\label{vr}\end{equation}
where the modification factor $W$ is defined as
\begin{eqnarray}
W(y) &=& 1+\frac{2+y^3Z(y)-(y^2-y+2)e^{-y}}{3y^2},\nonumber\\
Z(y) &=& \int_y^{\infty} {dt\over t} e^{-t}.
\end{eqnarray}
We emphasize that the general potential $V({\bf r},{\bf L})$ can be simplified as $V(r,L)$ only in the sense of the screening radius (\ref{eqrd}). For a more detailed calculation about a general potential, one may refer to the Refs.\cite{Mustafa:2004hf, Chakraborty:2006md} based on the linear response theory.

Now we apply the above transport solutions to the quarkonium
dissociation in hot and dense matter created in high-energy nuclear
collisions. The interaction between two quarks in vacuum can be well
characterized by the Cornell potential~\cite{Satz:2005hx}
$V_0(r)=-\alpha/r+\sigma r$ with coupling constant $\alpha=\pi/12$
and string tension $\sigma=0.2$ GeV$^2$. At finite temperature, the
screening potential for a stationary pair of heavy quarks can be
written as~\cite{Dixit:1989vq,Satz:2005hx}
\begin{equation}
\label{lattice}
V_0(r)=-{\alpha\over r}e^{-\mu r}-{\sigma\over
2^{3\over 4}\Gamma({3\over 4})}\left({r\over
\mu}\right)^{1/2}K_{1\over 4}\left((\mu r)^2\right),
\end{equation}
where $\Gamma$ and $K$ are the Gamma and modified Bessel functions.
The temperature of the medium is hidden in the screening mass
$\mu(T)$ which can be extracted~\cite{Satz:2005hx} from lattice QCD
calculated free energy~\cite{Kaczmarek:2004gv, Digal:2005ht}.

To establish a unique mapping between the relaxation length and
the velocity, we estimate the relaxation time of the hot and
dense matter by considering its electric analogue. When an electric
charge is put into a conducting medium, the medium is neutralized in
a time scale of $\tau =1/\left(4\pi\sigma_e\alpha_e\right)$,
where $\sigma_e$ is the electric conductivity of the medium, and
$\alpha_e$ is the fine-structure constant. We replace $\sigma_e$ by
the conductivity $\sigma_s\approx 0.4 T$ for a strong field, estimated from hot quenched lattice QCD
\cite{Aarts:2007wj}, and $\alpha_e$ by $\alpha$, the relaxation length becomes $L =15\upsilon/(2\pi^2T)$.

\begin{figure}[!hbt]
\centering
\includegraphics[width=0.7\textwidth]{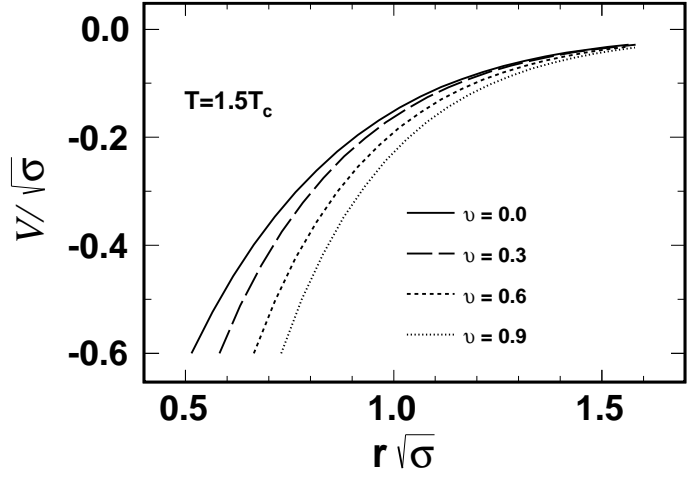}
\caption{ The velocity dependence of the screening potential at a fixed temperature $T=1.5T_c$. $\sigma$ is the string tension and the stationary potential is taken as the free energy~\cite{Kaczmarek:2004gv,
Digal:2005ht, Satz:2005hx}.}\label{fig2}
\end{figure}

In Fig.\ref{fig2} one sees the velocity induced change in the heavy
quark potential at a fixed temperature $T=1.5T_c$. The stationary
potential is taken from the lattice simulation Eq.~(\ref{lattice}). Since the screening length is proportional to the
velocity and inversely proportional to the temperature of the
medium, the potential well becomes deeper and screening becomes less effective, when the quarkonium velocity relative to the medium increases.

\begin{figure}[!hbt]
\centering
\includegraphics[width=0.7\textwidth]{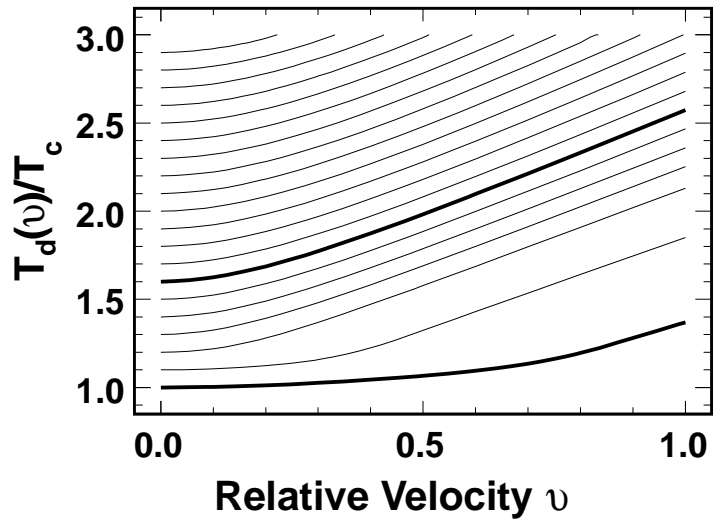}
\caption{ The scaled dissociation temperature $T_d(\upsilon)$, starting at different stationary values $T_d(0)$. $T_c=165$ MeV is the critical temperature of the quark matter, and the upper and lower thick lines are respectively for $J/\psi$ and the excited states $\psi'$ and $\chi_c$. }
\label{fig3}
\end{figure}

With the known potentials $V_0(T,r)$ and $V(T,r,L)$, the screen radius
$r_d(T,L)$ at finite temperature $T$\ can be calculated through
(\ref{eqrd}), where the temperature $T$\ dependent inherited from potential $V$\ is written explicitly. In the rest frame of the di-quark system, the
condition for dissociating a quarkonium should not depend on its
relative velocity, namely the critical screening radius is a
constant,
\begin{equation}
r_d(T_d,L(\upsilon, T_d))=C,
\label{eq_tdv}
\end{equation}
where the constant C can be calculated directly at $\upsilon=0$,
\begin{equation}
C=r_d(T_d(\upsilon=0),L=0)={1\over \mu}{1+{\pi\over 16\Gamma^2(3/4)}{\sigma\over \alpha\mu^2}\over 1+{1\over 4}{\sigma\over \alpha\mu^2}}
\end{equation}
with $\mu$\ the screening mass at $T_d(\upsilon=0)$. Thus when the dissociation temperature of a $J/\psi$\ at rest is given, the constant $C$\ can be calculated and Eq. (\ref{eq_tdv}) determines the dissociation temperature $T_d(\upsilon)$\ for a moving quarkonium with velocity $\upsilon$. When $T_d(0)$ runs from $T_c$ to
$2.5 T_c$, the screening radius $r_d(T_d(0),0)$ runs from $0.44$ fm
to $0.21$ fm. The velocity-dependent temperature $T_d(\upsilon)$ is shown
in Fig.\ref{fig3}. Since the lattice calculation of the stationary dissociation temperature $T_d(0)$ is still with some uncertainty, we take $T_d(0)$ as an adjustable parameter in Fig.\ref{fig3}. Once we fix $T_d(0)$, its velocity dependence can
be obtained from the corresponding curve. As expected, when a
quarkonium moves at a large velocity relative to the medium, the
screening effect becomes weaker and the dissociation temperature
becomes higher. The velocity induced shift of the dissociation
temperature can be as large as $\Delta T_d(\upsilon)\sim T_c$ for fast
quarkonia. For charmonium, the dissociation temperature is $T_d
\sim 1\textrm{-}2T_c$ at $\upsilon=0$~\cite{Satz:2005hx, Wong:2004zr} but goes up to
$1.2\textrm{-}2.7T_c$ at $\upsilon=0.8c$, see Fig.\ref{fig3}. Considering the fact that
the fireball temperature formed in heavy ion collisions at RHIC
energy is in the region $T\sim 1\textrm{-}2T_c$, the $J/\psi$ transverse
momentum spectrum should be sensitive to the velocity dependence of
the dissociation temperature. The much higher dissociation
temperature for fast moving $J/\psi$s will lead to a weaker
suppression in the high transverse momentum region.

In order to quantitatively see the effect of the velocity-dependent
temperature $T_d(\upsilon)$ on quarkonium suppression in high-energy nuclear
collisions, we take a detailed transport approach~\cite{Yan:2006ve}
to describe the dynamical evolution of the hot and dense medium. The
model contains transport equations for the quarkonium motion in the
medium and hydrodynamic equations for the space-time evolution of
the medium. The initial distribution of energy density and entropy density is based on Glauber Model. Both local temperature $T({\bf x}, t)$\ and local velocity $u({\bf x}, t)$\ that used in the transport eqaution for quarkonia are solved from the hydrodynamic equations as in our previous work~\cite{Liu:2009nb}. In heavy ion collisions there are two sources for quarkonium
production: the primordial production at the
initial state and the regeneration in the hot medium. During the
evolution, all of the produced quarkonia suffer from the medium induced dissociation, dominantly by the gluon interactions.
The model used here describes well both
$J/\psi$~\cite{Zhu:2004nw, Liu:2009wza} and
$\Upsilon$~\cite{Liu:2010ej} suppression. In order to demonstrate
the velocity effect on the quarkonia suppression, it is necessary to
study the transverse momentum distributions. We consider the differential nuclear modification factor
$R_{AA}(p_t)=N_{{AA}}(p_t)/\left(N_{{coll}}N_{{pp}}(p_t)\right)$ as
a function of quarkonium transverse momentum $p_t$. $N_{{AA}}(p_t)$
and $N_{{pp}}(p_t)$ are differential quarkonium yields in heavy ion
and elementary p+p collisions, and $N_{{coll}}$ is the number of
nucleon+nucleon collisions in heavy ion collisions. We will focus on
the high $p_t$ behavior of $R_{AA}(p_t)$.

\begin{figure}[!hbt]
\centering
\includegraphics[width=0.7\textwidth]{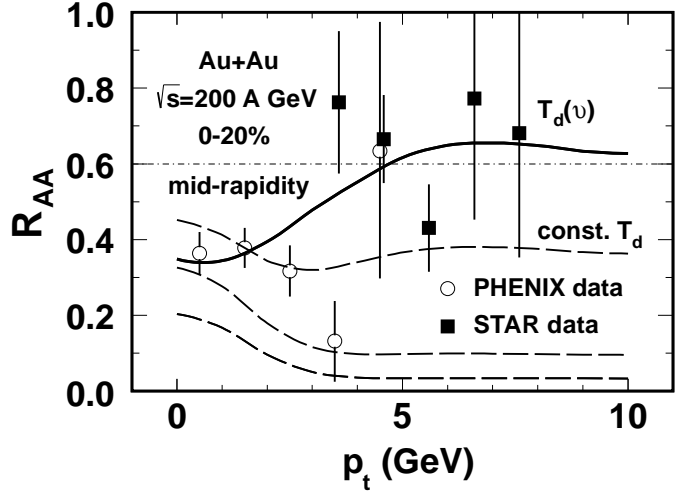}
\caption{ The $J/\psi$ nuclear modification factor $R_{AA}(p_t)$ at RHIC. The
data are from the PHENIX~\cite{Adare:2006ns} at rapidity $|y|<0.35$ and
STAR~\cite{Tang:2011kr} at rapidity $|y|<0.9$, the solid line is the
calculation with a velocity-dependent temperature starting at $T_d(0)/T_c=1.6$, and the dashed lines
are the calculations with a constant dissociation temperature $T_d=1.3T_c, 1.6T_c$ and $1.9T_c$ from bottom to top. }
\label{fig4}\end{figure}

Fig.\ref{fig4} shows the $J/\psi$ $R_{AA}(p_t)$ for a constant and a
velocity-dependent dissociation temperature in central Au+Au collisions at top
RHIC energy $\sqrt{s_{NN}}=200$ GeV. As one can see, at the low
$p_t$ region ($ \le 3.5$ GeV), the experimental results of $J/\psi$
$R_{AA}(p_t)$ is less than 0.4. However, in the higher $p_t$ region,
the value of the nuclear modification factor becomes higher
$R_{AA}(p_t) \approx 0.6$ indicating weaker suppression in $J/\psi$
yield. Note that this $p_t$ dependence can not be reproduced by
a constant dissociation temperature unless a strong Cronin effect is assumed
even at extremely high $p_t$~\cite{Liu:2009nb}. As discussed in
\cite{Adare:2012qf}, the strong Cronin effect at high $p_t$ region
is not favored by the latest $J/\psi$ data from d+Au collisions. In
our calculation, the Cronin effect has been characterized by a Gaussian
smearing scheme~\cite{Zhao:2007hh}, which contribute little to the
high $p_t$ region $p_t>6$ GeV. For the stationary charmonia, the
dissociation temperature calculated from the Schr\"odinger equation
is in between $(1.1 \textrm{-} 2.1) T_c$ for the ground state $J/\psi$ and
$~T_c$ for the excited states $\psi'$ and $\chi_c$, depending on the
used heavy quark potential~\cite{Digal:2001ue,Wong:2006dz, Satz:2005hx}.

Since the fireball temperature in a central collision is much higher
than $T_c$, almost all the excited states are dissociated in the
medium, we will consider mainly the ground state. Considering the
fact that the contribution from the decay of the excited states to
the final $J/\psi$s is about $40\%$, there is an upper limit of $0.6$\ for
$J/\psi\ R_{AA}$. In Fig.\ref{fig4} the three dashed lines represent
the results with a constant dissociation temperature $T_d = $1.3,
1.6, and 1.9$T_c$ from bottom to top, respectively. As one can see
in the figure, the numerical results with a constant $T_d$ are all
much less than $0.6$ and overestimate the $J/\psi$\ suppression.

We now analyze the results with the velocity-dependent
temperature, see the solid line in Fig.\ref{fig4}. Fitting the
experimental data of the nuclear modification factor $R_{AA}(N_{\textrm{part}})$ as a function of the number of participant nucleons $N_{\textrm{part}}$~\cite{Adare:2006ns}, we obtain $T_d(0)=1.6 T_c$. Note that for an expanding
fireball, $p_t=0$ in the laboratory frame corresponds generally to a
nonzero velocity in the rest frame of the fireball, therefore the
velocity-dependent temperature even at $p_t=0$ is already affected
by the velocity $v$. That is why the $R_{AA}$ at $p_t=0$ does not
coincide with the calculation with a constant $T_d=1.6T_c$. From
Fig.\ref{fig3}, one sees that the increase of the dissociation temperature is approximately linear at high velocity.  On the other hand, the maximum temperature of the fireball in a central Au+Au
collision is $T_{\textrm{max}}\approx2T_c$~\cite{Kolb:2001qz, Heinz:2001xi}.
Therefore, at sufficiently high transverse momentum, the dissociation
temperature may stay above the maximum temperature. For example, the velocity of those $J/\psi$s at $p_t\sim 5$\ GeV is above $0.8c$, and the dissociation temperature $T_d$\ is about $2.3T_c$. As a result,
those high $p_t$ $J/\psi$s will survive in the quark gluon plasma.
For the same reason, those high $p_t$ excited states $ \psi'$s and
$\chi_c$s produced in the more peripheral region where the
temperature is lower than their dissociation temperature will also
survive. The competition between the velocity ($p_t$) dependent
dissociation temperature and the fireball temperature leads to
$R_{AA}>0.6$ at high $p_t$, as shown in Fig.\ref{fig4}. It is clear
in the figure that our calculation with the velocity-dependent
temperature is consistent with the experimental observation. High
statistics data are needed in order to confirm this ansatz for
quarkonium suppression in high-energy nuclear collisions.

\begin{figure}[!hbt]
\centering
\includegraphics[width=0.7\textwidth]{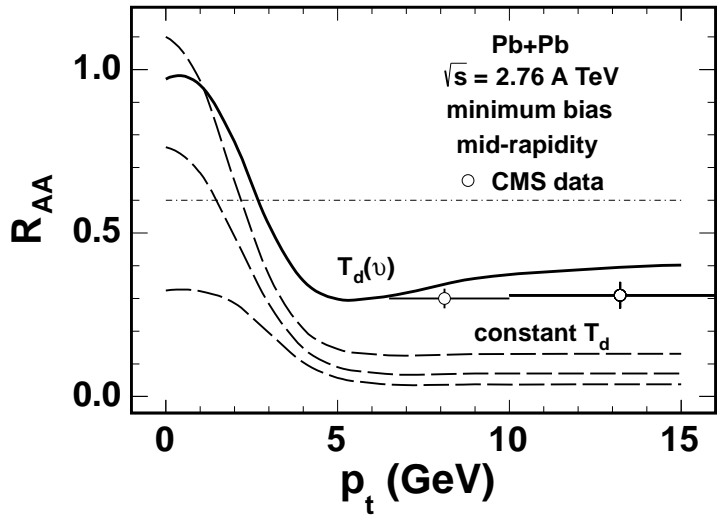}
\caption{ The $J/\psi$ nuclear modification factor $R_{AA}(p_t)$ at LHC. The data are
from the CMS Collaboration at rapidity $|y|<2.4$~\cite{Chatrchyan:2012np}, the solid line is the
calculation with a velocity-dependent temperature starting at $T_d(0)/T_c=1.6$, and the dashed lines
are the calculations with a constant dissociation temperature $T_d=1.3T_c, 1.6T_c$ and $1.9T_c$ from bottom to top. }
\label{fig5}\end{figure}
In order to further test the model, the $p_t$ dependence of the
$R_{AA}(p_t)$ for prompt $J/\psi$s from minimum bias Pb+Pb
collisions at LHC energy $\sqrt{s_{NN}}=2.76$ TeV is also calculated
in our model. The results are compared with the experimental
data~\cite{Chatrchyan:2012np} in Fig.\ref{fig5}. For this
calculation, the charm quark production cross section is taken as
$d\sigma_{NN}^c/dy= 0.62$ mb at midrapidity~\cite{Averbeck:2011ga,
ALICE:2011aa}. Significant regeneration has been reported with the
large charm cross section~\cite{Averbeck:2011ga, ALICE:2011aa}.
Since the charm quarks interact strongly with the medium, losing its
initial energy, the regenerated charmonia are soft, leading to a
large $R_{AA}$ in the low $p_t$ region. Similar to the case at RHIC,
the CMS experimental value of $R_{AA}\approx0.3$ at high $p_t$ can
be reproduced only when the velocity-dependent temperature is
considered, while all other results with a constant $T_d$
underpredict the values of $R_{AA}$. At the same $p_t \sim 6$ GeV
region, the value of $R_{AA}$ from LHC is much lower than that from
RHIC, implying that a much hotter medium has been formed in heavy
ion collisions at the higher energy. At $p_t\sim 10$\ GeV, the velocity is above $0.9c$, so that $T_d$\ is about $2.5T_c$, which is still smaller than the highest temperature of the fireball at LHC.
\par
In summary, we studied the heavy quark potential and dissociation
temperature for moving quarkonia in quark gluon plasma in
high-energy nuclear collisions. For a moving heavy quark pair in the
hot medium, the screening potential is reduced and the dissociation
temperature is enhanced. As a consequence of the velocity-dependent dissociation temperature, the $J/\psi$
suppression becomes significantly weaker at high transverse
momentum.

\appendix {\bf Acknowledgement:} The work is supported by the NSFC
under grant No. 11079024, the MOST under grant No.
2013CB922000, and the Helmholtz International Center for FAIR
within the framework of the LOEWE program launched
by the State of Hesse.

\end{document}